\documentclass[12pt]{article}

\usepackage{amssymb}
\usepackage[dvips]{graphicx}

\def\appendix#1{
  \addtocounter{section}{1}
  \setcounter{equation}{0}
  \renewcommand{\thesection}{\Alph{section}}
  \section*{Appendix \thesection\protect\indent \parbox[t]{11.715cm} {#1}}
  \addcontentsline{toc}{section}{Appendix \thesection\ \ \ #1}
  }

\newcommand {\bd}{\begin{displaymath}}
\newcommand {\ed}{\end{displaymath}}
\newcommand {\eq}{\begin{equation}}
\newcommand {\beq}{\begin{equation}}
\newcommand {\eeq}{\end{equation}}
\newcommand {\beqa}{\begin{eqnarray}}
\newcommand {\eeqa}{\end{eqnarray}}
\newcommand {\n}{\nonumber \\}
\newcommand {\tr}{{\rm tr\,}}
\newcommand {\Tr}{\mbox{Tr\,}}
\newcommand {\eff}{\mbox{\scriptsize eff}}

\newcommand {\Pf}{\mbox{Pf}}
\newcommand {\Imag}{\mbox{Im}}

\newcommand {\ee}{\mbox{e}}

\newcommand {\dd}{\mbox{d}}

\newcommand {\del}{\partial}

\font\mybb=msbm10 at 12pt
\def\bb#1{\hbox{\mybb#1}}

\def\IR{{\bb R}}

\begin{document}

\setlength{\oddsidemargin}{0cm}
\setlength{\baselineskip}{7mm}

\begin{titlepage}

\baselineskip=14pt

\renewcommand{\thefootnote}{\fnsymbol{footnote}}
\begin{normalsize}
\begin{flushright}
\begin{tabular}{l}
NBI-HE-00-15\\
hep-th/0003223\\
\hfill{ }\\
March 2000
\end{tabular}
\end{flushright}
  \end{normalsize}


\vskip 2 cm

\vspace{1cm}

\vspace*{0cm}
    \begin{Large}
       \begin{center}
{Spontaneous Breakdown of Lorentz Invariance in IIB Matrix Model}\\

       \end{center}
    \end{Large}
\vspace{1cm}

\begin{center}
           Jun N{\sc ishimura}\footnote{
Permanent address : Department of Physics, Nagoya University,
Nagoya 464-8602, Japan,\\
e-mail address : nisimura@nbi.dk}
\setcounter{footnote}{2}
           {\sc and}
           Graziano V{\sc ernizzi}\footnote{
e-mail address :
vernizzi@nbi.dk}\\
      \vspace{1cm}
         {\it The Niels Bohr Institute\\ Blegdamsvej 17, DK-2100
                 Copenhagen \O, Denmark}\\[4mm]
\end{center}

\vskip 2 cm

\hspace{5cm}

\begin{abstract}
\noindent
We study the IIB matrix model, 
which is conjectured to be a nonperturbative definition
of superstring theory,
by introducing an integer deformation parameter $\nu$
which couples to the 
imaginary part of the effective action induced by fermions.
The deformed IIB matrix model continues to 
be well-defined for arbitrary $\nu$,
and it preserves
gauge invariance, Lorentz invariance,
and the cluster property.
We study the model at $\nu=\infty$ using a saddle-point analysis, 
and show that ten-dimensional Lorentz invariance is spontaneously broken
at least down to an eight-dimensional one.
We argue that it is likely that 
the remaining eight-dimensional Lorentz invariance is 
further broken, which can be checked
by integrating over the saddle-point configurations 
using standard Monte Carlo simulation.
\end{abstract}
\vfill
\end{titlepage}
\vfil\eject
\setcounter{footnote}{0}

\setcounter{equation}{0}
\renewcommand{\thefootnote}{\arabic{footnote}}

\baselineskip=18pt

\setcounter{footnote}{0}

\renewcommand{\thefootnote}{\arabic{footnote}}
Recent developments in string dualities have suggested that the known
five types of superstring theories in ten dimensions and 
M-theory in eleven dimensions are just different microscopic descriptions
of the same underlying physics.
This means that if any of the descriptions can be defined nonperturbatively,
it is as good as any other
to study the 
dynamics of 
the universality class of string/M theories.
A celebrated first step towards this end has been made by 
Ref. \cite{BFSS}, where the so-called Matrix theory has been proposed as
a nonperturbative definition of M-theory in the infinite momentum frame.
Soon after, 
the IIB matrix model \cite{IKKT} has been proposed as a nonperturbative
definition of type IIB superstring theory.
Although there are many evidences that support these conjectures,
there is no direct proof that these models reproduce the string/M theories
perturbatively\footnote{See, however, Ref. \cite{FKKT}
as an attempt in this direction.}.
If these conjectures are true, 
there is a hope 
to understand all the fundamental questions 
about the Standard Model, 
including the space-time dimensionality, gauge group, matter contents, 
the hierarchy problem, 
the cosmological constant problem, and so on,
in terms of the dynamics of these models.



So far, the understanding of the vacuum of these models
is quite limited.
In this Letter, we attempt to extract information about
the dimensionality of the space-time which
is dynamically generated in the IIB matrix model.
The IIB matrix model is a supersymmetric matrix model obtained
formally by taking a zero-volume limit \cite{EK} of ten-dimensional
SU($N$) super Yang-Mills theory.
Unlike in field theories,
the space-time is treated in this model as a dynamical object,
which is represented by ten bosonic $N \times N$ hermitian matrices, 
where $N$ should be sent to infinity eventually.
Therefore, the model has a potential 
to explain the dynamical origin of the 
dimensionality of our space-time,
which, in the Standard Model, has to be given as an input parameter. 
%
If we are ever going to explain 
our four-dimensional space-time in this way,
the vacuum 
of the IIB matrix model should be
dominated by configurations 
which have only four-dimensional extent\footnote{This is similar
in spirit to a description of
our four-dimensional space-time as a ``brane'' in a higher-dimensional
{\em non-compact} space-time, which is proposed \cite{RS}
as an alternative to a more conventional
compactification approach.}.
This means, in particular, that the ten-dimensional Lorentz invariance
of the model should be spontaneously broken to a four-dimensional 
one.\footnote{The possibility of a non-trivial vacuum structure in 
string theory has been considered also in \cite{AK}.}


Such 
an
issue can 
{\em in principle} be addressed by performing Monte Carlo simulation,
since the model is completely well defined for arbitrary $N$
without any cutoff \cite{SuyamaTsuchiya,AIKKT,KNS}.
Indeed the large-$N$ dynamics of 
the four-dimensional version of the IIB matrix model, 
which can be obtained by a  zero-volume limit
of four-dimensional super Yang-Mills theory, 
has been understood to a considerable extent through Monte Carlo simulation 
\cite{AABHN}.
Comparison of the results with those for the corresponding 
bosonic theory \cite{HNT}, 
revealed that supersymmetry indeed plays an 
important role in the dynamics of large-$N$ reduced models
as a nonperturbative definition of string theories.
Unfortunately, such a direct approach does not work in the IIB matrix model
as it stands,
due to the fact that the effective action induced by fermions
is generically complex,
whereas in the above mentioned four-dimensional version, it is real.
In general, when the action of a theory has a non-zero imaginary part,
the number of configurations needed to extract any information
increases as exponential of the system size.
This is the notorious
sign problem, which occurs also
in many other interesting systems related to particle physics, 
such as theories with a chiral fermion (as in the present case),
theories with a $\theta$ vacuum, Chern-Simons theories
and theories with a finite baryon number density.
We emphasize, however, that the problem is a purely technical one,
and indeed its complete solution in some particular class
of systems has been obtained recently \cite{sign}.

%
In order to search for a way out of this difficulty,
let us consider a deformation of the IIB matrix model by
introducing
an integer parameter $\nu$, which couples
to the 
imaginary part of the effective action induced by fermions.
The deformed model, which reduces to the IIB matrix model
at $\nu=1$, 
is well-defined for arbitrary $\nu$.
It preserves both gauge invariance and Lorentz invariance,
and 
%
moreover, it preserves the cluster property, which is an important 
consequence of the supersymmetry of the original model.
At $\nu=0$, the sign problem disappears and standard Monte Carlo
techniques become applicable.
Monte Carlo studies up to $N=512$ show, however, that
the ten-dimensional Lorentz invariance is 
{\em not} spontaneously broken \cite{oneloop},
which suggests that if Lorentz invariance is spontaneously
broken in the original model,
the imaginary part of the effective action must play a 
crucial role.
This motivated us to consider the opposite extreme,
namely $\nu = \infty$.

At $\nu = \infty$, 
the integration over the bosonic matrices
is dominated by the
saddle-points of the imaginary part of the effective action.
We find that the saddle-points are
given by configurations 
which have only eight-dimensional extent.
This implies
that 
the ten-dimensional Lorentz invariance {\em is} 
spontaneously broken 
at least to an eight-dimensional one.
Since we still have to integrate over 
the saddle-point configurations,
the question arises whether 
the remaining eight-dimensional Lorentz invariance further breaks down, 
say, to a four-dimensional one.
To this end, we study the hessian at configurations
which have only $d$-dimensional extent ($d\le 8$)
and find that it is zero of order $(8-d)\, \{2\, (N^2-1)-16\}$.
In other words, the imaginary part of the
effective action becomes {\em more} stationary for smaller $d$.
This gives a huge enhancement
to configurations with smaller $d$,
which is found to cancel exactly
the entropical barrier against having such configurations.
%
Considering the effect of the real part of the effective action,
which is studied in Ref. \cite{AIKKT} using perturbation theory, 
it is plausible that the vacuum of the model at $\nu = \infty$
is actually given by $d<8$.
We also find a lower bound on $d$ as $d > 2$.
Further information about the vacuum can be obtained
by performing explicitly
the integration over the 
saddle-point configurations by Monte Carlo simulation.
Naively, one might consider that the sign problem, which already 
exists in the original model ($\nu = 1$), 
becomes maximally severe at $\nu = \infty$.
This is not the case, as we will see.
%




\vspace*{1cm}


The IIB matrix model is formally a zero-volume limit of
ten-dimensional pure ${\cal N}=1$ supersymmetric Yang-Mills theory.
The action, therefore, is given by
\beqa
S &=& S_b + S_f \ ,\n
S_b &=& -\frac{1}{4} \, \tr \left( [A_{\mu},A_{\nu}]^2 \right) \ , \n
S_f  &=& - \frac{1}{2} \,
\tr \left( \psi _\alpha (\, {\cal C} \,  \Gamma_{\mu})_{\alpha\beta} 
[A_{\mu},\psi _\beta] \right)  \ .
\label{action}
\eeqa
$A_\mu$ ($\mu = 1,\cdots,10$) and 
$\psi_\alpha$ ($\alpha = 1,\cdots , 16$) are
$N \times N$ traceless hermitian matrices,
which can be expanded in terms of the generators 
$t^a$ of SU($N$) as
\beq
(A_{\mu })_{ij} = \sum_{a=1}^{N^2-1} 
A_{\mu}^{ a} \, (t^{a})_{ij} ~~~~~;~~~~~  
(\psi_\alpha ) _{ij} = \sum_{a=1}^{N^2-1} 
\psi_{\alpha}^{ a} \, (t^{a})_{ij} \ , 
\eeq
where $A_\mu ^{a}$ is a real variable and 
$\psi_{\alpha}^{a}$ is a real Grassmann variable.
We have made a Wick rotation in the action (\ref{action}),
so that the metric has Euclidean signature.
The $16 \times 16$ matrices $\Gamma _\mu$ are 
ten-dimensional gamma matrices after Weyl projection,
and the unitary matrix ${\cal C}$ is a 
charge conjugation matrix satisfying
\beq
{\cal C} \, \Gamma _\mu \, {\cal C}^\dag
= (\Gamma _\mu)^\top ~~~~~;~~~~~
{\cal C} ^\top = {\cal C}  \ .
\label{chargeconj}
\eeq
The model has a manifest ten-dimensional 
Lorentz invariance,
by which we actually mean an SO(10) invariance.
$A_{\mu}$ transforms as a 
vector and
$\psi _\alpha$ transforms as a Majorana-Weyl spinor.
%


Pure ${\cal N}=1$ supersymmetric Yang-Mills theory
can be also defined in 3D, 4D and 6D, as well as in 10D.
Hence, by taking a zero-volume limit of these theories,
we can define supersymmetric large-$N$ reduced models,
which are $D=3,4,6$ versions of the IIB matrix model.
A nontrivial question then concerns
whether the integration over the bosonic matrices 
is convergent, since 
the integration domain for hermitian matrices is non-compact.
A potential danger of divergence, 
even for finite $N$, exists when 
the eigenvalues of $A_\mu$ become large. 
This issue has been addressed in Ref. \cite{AIKKT}
using a one-loop perturbative argument.
When all the eigenvalues are well separated from each other,
one can expand the matrices $A_\mu$ and $\psi _\alpha$ 
around diagonal matrices as
\beq
(A_\mu)_{ij} = x_{i \mu} \, \delta _{ij} + a_{\mu ij} ~~~~~;~~~~~
(\psi_\alpha)_{ij}  = \xi_{i \alpha} \, \delta _{ij} + \varphi_{\alpha ij}  \ ,
\label{perturbation}
\eeq
where $a_\mu$ and $\varphi_\alpha$ are matrices containing only off-diagonal
elements.
One can then integrate over the off-diagonal elements
$a_\mu$ and $\varphi_\alpha$ up to one-loop,
by fixing the gauge properly and including the Faddeev-Popov ghosts.
The integration over the bosonic off-diagonal elements $a_\mu$ 
and the Faddeev-Popov ghosts, 
gives a logarithmic attractive potential between
all the pairs of $x_i$ \cite{BHN}.
This potential, however, is exactly cancelled by the contribution of the
fermionic off-diagonal elements $\varphi_\alpha$ due to supersymmetry.
This cancellation is responsible for the cluster property of the model
\cite{IKKT}, which is important for the interpretation of the model
as a string theory.
In order to calculate the effective potential for $x_{i \mu}$,
one still has to integrate 
over the fermionic diagonal elements $\xi_{i \alpha}$, 
which is nontrivial.
In Ref. \cite{AIKKT}, it has been shown that
the effective potential for $x_{i\mu}$ can be given by
a branched-polymer like attractive interaction among $N$ points in
$D$-dimensional space-time represented by $x_{i\mu}$ ($i=1,\cdots ,N$).
Now counting the power of $x_i$ in the partition function, 
the integration measure for $x_{i \mu}$
gives $D\, (N-1)$ and 
the integration over $\xi _{i \alpha}$ gives
$- \frac{3}{2}\, p \, (N-1)$, where $p=2\, (D-2)$ is the number of
real components of the spinor considered for each $D$.
Therefore, naively one concludes that when 
\beq
D\, (N-1) - \frac{3}{2} \, p \, (N-1) = 2 \, (3-D) \, (N-1) \ge 0  \ , 
\eeq
the integration over $x_{i \mu}$ is divergent
and otherwise convergent.
This means that the integral is divergent for $D=3$ and 
convergent for $D=4,6,10$, irrespectively of $N$.
This conclusion is in agreement with
an exact result available for $N=2$ \cite{SuyamaTsuchiya}
and a numerical result obtained for $N=3$ \cite{KNS}.
For $D=4$, 
Monte Carlo simulations show that the conclusion extends
to the large-$N$ limit \cite{AABHN}.
Therefore, 
it is conceivable
that the above conclusion obtained by the 
one-loop argument holds in general, 
although it is not a rigorous proof.


Going back to the definition of the model (\ref{action}),
let us first integrate over the fermionic matrices $\psi _ \alpha$,
which induces an effective action for $A_\mu$.
When fermions are real Grassmann variables as in the present $D=10$ case,
the fermion integral yields
a pfaffian of a $p \, (N^2 -1) \times p \, (N^2 -1)$ matrix.
On the other hand, when fermions are complex Grassmann variables 
as in the $D=4,6$ case,
the fermion integral yields
a determinant of a 
$p \,(N^2 -1)/2 \times p \, (N^2 -1)/2$ matrix.
For $D=4$, the fermion determinant is real positive, which allows
a direct Monte Carlo study of the model \cite{AABHN}.
For $D=6,10$, the fermion determinant ($D=6$) or pfaffian ($D=10$)
is complex in general, since fermions are essentially chiral in these
cases. This causes the notorious sign problem.
There are some exceptions\footnote{This is because,
when $(N^2 -1)$ is smaller than $D$, one can always make a 
$D$-dimensional rotation to set $A_{\mu} = 0$ for
$\mu = 1, \cdots , D-(N^2 -1)$,
since one can regard $A_{\mu}^a$ as $(N^2 -1)$, $D$-dimensional 
real vectors.
The statements in $D=10$ for $N=2$ and $N=3$, for example, follow from
(\ref{omega6}) and (\ref{omega9}), respectively.} 
when $N$ is small \cite{SuyamaTsuchiya,KNS}.
For $D=6$, the fermion determinant is real positive for $N=2$.
For $D=10$, the pfaffian is real positive for $N=2$ and
real (but not positive definite) for $N=3$.
Although we consider only the $D=10$ case
in what follows, 
the analysis can be readily applied to the $D=6$ case as well.
We also restrict ourselves to $N \ge 4$ so that 
the pfaffian is complex generically.

In $D=10$, the fermion integral yields
\beq
Z_f [A] = 
 \int \dd \psi   ~ \ee ^{-S_f } = \Pf \, {\cal M}  \ ,
\eeq
where
\beq
{\cal M}_{a \alpha ,  b \beta} = 
-  i \, f_{abc} (\,{\cal C} \, \Gamma _\mu)_{\alpha \beta} A_\mu ^c  
\eeq
is a $16 \,  (N^2-1) \times 16 \, (N^2-1)$ anti-symmetric matrix, regarding
each of $(a\alpha)$ and $(b\beta)$ as a single index.
The real totally-antisymmetric tensor $f_{abc}$ gives 
the structure constants of SU($N$).
In what follows, it proves convenient to work with an explicit 
representation of the gamma matrices given by
\beqa
&~&\Gamma _1 = i \, \sigma_2  \otimes \sigma_2  
              \otimes \sigma_2  \otimes \sigma_2 ~;~
\Gamma _2 = i \, \sigma_2  \otimes \sigma_2
              \otimes {\bf 1} \otimes \sigma_1 ~;~
\Gamma _3 = i \, \sigma_2  \otimes \sigma_2 
              \otimes {\bf 1} \otimes \sigma_3 ~;~\n
&~&\Gamma _4 = i \, \sigma_2  \otimes \sigma_1
              \otimes \sigma_2 \otimes {\bf 1} ~;~
\Gamma _5 = i \, \sigma_2  \otimes \sigma_3
             \otimes \sigma_2 \otimes {\bf 1} ~;~
\Gamma _6 = i \, \sigma_2  \otimes {\bf 1}
              \otimes \sigma_1 \otimes \sigma_2  ~;~ \n
&~&\Gamma _7 = i \, \sigma_2  \otimes {\bf 1}
             \otimes \sigma_3 \otimes \sigma_2 ~;~
\Gamma _8 = i \, \sigma_1  \otimes {\bf 1}
              \otimes {\bf 1} \otimes {\bf 1} ~;~
\Gamma _9 = i \, \sigma_3  \otimes {\bf 1}
              \otimes {\bf 1} \otimes {\bf 1} ~;~\n
&~&\Gamma _{10} = {\bf 1}  \otimes {\bf 1} \otimes {\bf 1} \otimes {\bf 1} \ ,
\label{Gamma}
\eeqa
for which the charge conjugation matrix ${\cal C}$ becomes a unit matrix.

Some important properties of the pfaffian $\Pf \, {\cal M}(A)$
are in order. First of all, it is invariant under a ten-dimensional 
Lorentz transformation,
$A'_\mu = \Lambda _{\mu \nu} A_\nu$, 
where $\Lambda$ is an SO(10) matrix.
It is also invariant under a gauge transformation
$A'_\mu = g   A_\nu g ^\dag$,
where $g$ is an SU($N$) matrix.
Under a parity transformation,
\beq
A^P _{10}  = - A_{10} ~~~~~;~~~~~
A^P_i  =   A_i ~~~~~(\mbox{for}~~~i=1,\cdots, 9) \ ,
\label{parity}
\eeq
the pfaffian becomes complex conjugate,
\beq
\Pf \, {\cal M}(A^P)  = \{ \Pf \, {\cal M}(A)\} ^*  \ ,
\label{paritytransf}
\eeq
since ${\cal M} (A^P) = {\cal M} (A) ^*$.
As a consequence, the pfaffian $\Pf \, {\cal M}(A)$ is real,
when $A_{10}=0$.
This also means that 
$\det {\cal M} = (\Pf \, {\cal M})^2$ is real positive,
when $A_{10}=0$.


We next consider the case with $A_9 =A_{10} = 0$.
In this case, the pfaffian is actually equal to
the fermion determinant 
of eight-dimensional Weyl fermion
which we denote as $\det {\cal M}^{(8)}$.
${\cal M}^{(8)}$ is an $8 \,(N^2-1)\times 8\, (N^2-1)$ matrix 
defined as ${\cal M}^{(8)}_{a \alpha ,  b \beta} = 
-  i \, f_{abc} (\tilde{\Gamma} _\mu)_{\alpha \beta} A_\mu ^c $,
where the eight-dimensional gamma-matrices $\tilde{\Gamma}_\mu$
after Weyl projection are given, for example, by
\beqa
&~& \tilde{\Gamma} _1 = i \, \sigma_2 \otimes \sigma_2  \otimes \sigma_2 ~;~
\tilde{\Gamma} _2 = i \, \sigma_2 \otimes {\bf 1} \otimes \sigma_1 ~;~
\tilde{\Gamma} _3 = i \, \sigma_2 \otimes {\bf 1} \otimes \sigma_3 ~;~
\tilde{\Gamma} _4 = i \,  \sigma_1 \otimes \sigma_2 \otimes {\bf 1} ~;~\n
&~&\tilde{\Gamma} _5 = i \, \sigma_3  \otimes \sigma_2 \otimes {\bf 1} ~;~
\tilde{\Gamma} _6 = i \,  {\bf 1} \otimes \sigma_1 \otimes \sigma_2  ~;~
\tilde{\Gamma} _7 = i \,  {\bf 1}  \otimes \sigma_3 \otimes \sigma_2 ~;~
\tilde{\Gamma} _8 =   {\bf 1} \otimes {\bf 1} \otimes {\bf 1} \ .
\label{Gamma8}
\eeqa
With this representation, one finds that ${\cal M}^{(8)}$
is a real matrix, which means that
$\det {\cal M}^{(8)} $ is real.
However, it is not positive definite, as we have checked
numerically.
In fact, this is the case
even for $A_8 = 0$.
However, if we take $A_7 = A_8 = 0$,
it becomes positive definite,
since in this case, we have
$\det {\cal M}^{(8)} = | \det {\cal M}^{(6)}| ^2  $,
where $\det {\cal M}^{(6)}$ is the fermion determinant 
of six-dimensional Weyl fermion \cite{KNS}.

Finally, when $A_3 =A_4 = \cdots = A_{10} = 0$,
one finds that $\Pf \, {\cal M} (A)=0$ \cite{KNS}.
This can be proved in the following way.
First we note that 
$(\Pf \, {\cal M})^2  =\det {\cal M} = |\det U|^{16}$, where
$U$ is an $(N^2 -1)\times (N^2 -1)$ matrix defined as
$U_{ab}=f_{abc} X^c$, where $X^c = A_1^c + i \, A_2^c$.
Since $U_{ab}X^b=0$, the matrix $U$ has a zero-eigenvalue,
and therefore $\det  U = 0$.
This completes the proof. 

Using Lorentz invariance of the pfaffian $\Pf \, {\cal M}(A)$,
we summarize some of its important properties as follows.
Let us first define sets of ``degenerate'' 
configurations $\Omega _d$ as
\beq
\Omega _d = \{ ~~\{ A_\mu\}~~ ;~~ n_\mu ^{(i)} A_\mu = 0 ~~\mbox{for}~~
\exists \, n_\mu ^{(i)} (i=1,\cdots , 10-d)~~\mbox{linearly independent}\} \ .
\eeq
Obviously, $\Omega _1 \subset \Omega _2 \subset 
\cdots \subset \Omega _9$.
Then the statements are
\beqa
&~&\mbox{(a)~When}~\{A_\mu \}\in \Omega _9 \ , ~~~\Pf \, {\cal M}(A)\in \IR \ .
\label{omega9}
\\
&~&
\mbox{(b)~When}~\{A_\mu \}\in \Omega _6 \ , ~~~\Pf \, {\cal M}(A) \ge 0 \ .
\label{omega6}
\\
&~& \mbox{(c)~When}~\{A_\mu \}\in \Omega _2 \ , ~~~\Pf \, {\cal M}(A) = 0 \ .
\label{omega2}
\eeqa
Here we recall that in the IIB matrix model, 
the space-time is treated as a
dynamical object represented by the bosonic matrices $A_\mu$.
In this regard,
generic configurations in $\Omega _d$ 
describe $d$-dimensional space-time.
It is very suggestive
that the phase of the pfaffian is sensitive to the
dimensionality $d$ of the space-time.
%
Indeed, the above results will play a crucial role
in the following analysis.

\vspace*{1cm}


Let us write the effective action induced by fermion integral
as $\Gamma_{\eff}
= - \ln (\Pf \, {\cal M})$, which we decompose into the real part
$\Gamma ^{(r)}$ and the imaginary part $\Gamma ^{(i)}$ as
$\Gamma _{\eff} =\Gamma ^{(r)} + i \, \Gamma ^{(i)}$.
Then we generalize the theory by introducing an integer parameter $\nu$,
which couples only to $\Gamma ^{(i)}$, the imaginary part of the 
effective action, as 
\beq
Z_{\nu} = 
\int \dd A  ~ \ee ^{-S_b[A] - \Gamma ^{(r)} [A] } 
~ \ee ^{- i \, \nu \, \Gamma^{(i)} [A]} \ .
\label{fullPTnu}
\eeq
The parameter $\nu$ has to be 
an integer since $\Gamma^{(i)}$ is defined only up to
modulo $2 \pi$.
For $\nu =1$, the model reduces to the original model.
Since $\Gamma^{(i)}$ flips its sign under a parity transformation
due to (\ref{paritytransf}),
the models with $\nu$ and $-\nu$ are nothing but parity partners.
Since $\Gamma ^{(r)}$ and $\Gamma ^{(i)}$ are both Lorentz invariant
and gauge invariant separately, the generalized model is also
both Lorentz invariant and gauge invariant.
According to 
the one-loop perturbative argument given below (\ref{perturbation}),
the deformed IIB matrix model continues to be
well-defined for arbitrary $\nu$, and moreover, it preserves the
cluster property, which is an important consequence of the
supersymmetry of the original model.

Another important point about the generalization we consider is that
$\langle  S_b
\rangle$ can be calculated
exactly and the result is independent of the parameter $\nu$.
For this, we recall that
in the original supersymmetric large-$N$ reduced models ($D=4,6,10$),
one obtains
\beq
\langle  
S_b \rangle
= \frac{1}{4} \, \left(D + \frac{p}{2}\right) (N^2 -1 ) 
= \frac{1}{2} \,  (D-1) \, (N^2 -1 )  \ ,
\label{trF2exact}
\eeq
using a scaling argument similar to \cite{HNT}.
Due to the fact that $\Gamma ^ {(i)} [A]$ is invariant under
a scale transformation of $A_\mu$,
one can easily find that the above result is unaltered by
the generalization to arbitrary $\nu$ for $D=6,10$.
In particular, this holds true even in the $\nu \rightarrow \infty$ limit.
%
Note, however, that this does not necessarily imply
that all the vacuum expectation values of the generalized model
are independent of $\nu$.
Even their convergence in the $\nu \rightarrow \infty$ limit 
is nontrivial.
This point shall be clarified later.

We prove a property of $\Gamma ^{(i)}$, which
turns out to be essential in the analysis of 
the $\nu \rightarrow \infty$ limit.
We first recall that
the pfaffian $\Pf \, {\cal M}(A)$ is
a polynomial of $A_\mu ^{a}$ of order $8 \, (N^2 -1)$.
Hence, $\Gamma ^{(i)}$ is
infinitely differentiable at a configuration,
for which the matrix ${\cal M}$ is invertible.
When the configuration belongs to $\Omega_d$ ($d=1,\cdots , 8$),
we find that 
\beq
\frac{\del ^n \, \Gamma ^{(i)}}
{\del A_{\mu_1} ^{a_1} \del A_{\mu_2} ^{a_2}
\cdots \del A_{\mu_n} ^{a_n} } 
=
%
- \frac{1}{2} \, 
\frac{\del ^n }{
\del A_{\mu_1} ^{a_1} \del A_{\mu_2} ^{a_2}
\cdots \del A_{\mu_n} ^{a_n}}
\,  \Imag \, \ln \det {\cal M } 
= 0 ~~~~~\mbox{for}~n=1,\cdots , (9-d) \ .
\label{derivatives}
\eeq
This is because, 
up to $(9-d)$-th order of perturbations, the configuration stays
within $\Omega _9$, and therefore $\det {\cal M}$ 
remains to be real positive due to (\ref{omega9}).

In the $\nu \rightarrow \infty$ limit,
the integration over $A_\mu$ is dominated by the configurations
which satisfy the saddle-point equation given by
\beq
0 = \frac{\del \, \Gamma ^{(i)}}{\del A_\mu ^a} =
- \frac{1}{2} \,
\frac{\del}{\del A_\mu ^a} \, \Imag \, \ln \det {\cal M}  
= - \frac{1}{2} \, \Imag \, \Tr \left({\cal M}^{-1} {\cal B} ^{(\mu a)} 
\right)  \ ,
\label{stationaryphase}
\eeq
where ${\cal B}^{(\mu a)} $ 
is an $A$-independent $16 \, (N^2-1) \times 16 \, (N^2-1)$ matrix defined by 
\beq
\left({\cal B}^{(\mu a)}\right)_{b \beta , c \gamma} 
= \frac{\del \, {\cal M}_{b\beta , c\gamma}}{\del A_\mu ^a}
= -i \, f_{abc} (\, {\cal C} \,  \Gamma_\mu)_{\beta\gamma}  \ .
\eeq
Since the equation (\ref{stationaryphase})
gives $10 \, (N^2 -1)$ constraints among
$10 \, (N^2 -1)$ real variables,
naively one would expect that 
only a very small portion of the configuration space survives.
This is not the case, however.
In fact, we find that
all the configurations that belong to $\Omega _8$
satisfy the saddle-point equation (\ref{stationaryphase})
due to (\ref{derivatives}).
%
The existence of solutions to (\ref{stationaryphase}) 
other than of this type, cannot be excluded.
It is reasonable, however, to consider
that the configurations in $\Omega _8$ dominate
in the sense of Lebesgue measure.
Hence,
at $\nu = \infty$
the full ten-dimensional Lorentz invariance is broken down at least to
an eight-dimensional one.
In order to examine whether the remaining eight-dimensional 
Lorentz invariance further breaks down,
we still have to integrate over the saddle-point configurations.
An important point here is that the imaginary part of the effective
action $\Gamma ^{(i)}$ for the saddle-point configurations
takes only 0 or $\pi$ due to (\ref{omega9}).
This means that there are actually two sequences of $\nu$,
(I)~$\nu = 0,2,4,\cdots,  2\infty$ and 
(II)~$\nu = 1,3,5,\cdots,  (2\infty +1)$,
which give two {\em a priori} different limiting theories.
%
%

In order to formulate the integration over
the saddle-point configurations,
we first rotate them
so that they satisfy $A_9 = A_{10}=0$
and then integrate over $A_1,A_2,\cdots, A_8$.
We define the Hesse matrix for $\Gamma ^{(i)}[A]$ 
within the directions transverse to the integration domain as
\beq
H_{ja,kb} =
\frac{\del ^2 \, \Gamma ^{(i)}}{\del A_j ^a \del A_k ^b}  =
%
- \frac{1}{2} \, \frac{\del ^2 }{\del A_j ^a
\del A_k ^b} \,  \Imag \, \ln \det {\cal M } 
=  
\frac{1}{2} \,  \Imag \, \Tr \left({\cal M}^{-1} {\cal B} ^{(j a)}
{\cal M}^{-1} {\cal B} ^{(k b)}  \right)  \ ,
\eeq
where $j,k = 9,10$.
We find that the Hesse matrix $H_{ja,kb}$ 
has $16$ zero-eigenvalues with eigenvectors corresponding to
perturbations 
\beq
(\delta A_9 ^a ,\delta A_{10}^a ) =
(A_1 ^a , 0),(A_2 ^a , 0), \cdots , (A_8 ^a , 0), 
 (0 , A_1 ^a ),(0 , A_2 ^a ), \cdots , (0 , A_8 ^a)    \ .
\label{zeromodes}
\eeq
These zero-modes are a reflection of the fact that
the configurations after these perturbations still stay within
$\Omega _8$, and thus satisfy
the saddle-point equation (\ref{stationaryphase}).
When $N$ is even, the Hesse matrix $H_{ja,kb}$ 
has actually two more zero-eigenvalues.
To see this,
we first note that $H_{9a,9b}=H_{10a,10b}=0$ and 
$H_{9a,10b}=H_{10b,9a}$.
Note also that $H_{9a,10b}=-H_{9b,10a}$
due to (\ref{paritytransf}),
which means that
an $(N^2 -1)\times (N^2 -1)$ matrix
$K_{ab}$ defined by $K_{ab} = H_{9a,10b}$
is antisymmetric.
Due to a general property of an antisymmetric matrix of odd size,
$K_{ab}$ for even $N$ should have a zero-eigenvalue, 
whose corresponding eigenvector we denote as $\chi ^a$.
Then one finds that 
$(\delta A_9 ^a ,\delta A_{10}^a ) = (\chi ^a , 0), (0,  \chi ^a )$
are eigenvectors of the Hesse matrix $H_{ja,kb}$ with zero-eigenvalues.
Unlike the 16 zero-modes in (\ref{zeromodes}),
these two additional zero-modes, which exist only for even $N$,
have nothing to do with the symmetry of the space of solutions
to the saddle-point equation (\ref{stationaryphase}).
Hence, they should be considered as accidental zero-modes.
Since the saddle-point analysis in such a case becomes more complicated,
we restrict ourselves to the odd $N$ case in what follows.

In order to deal with the 16 zero-modes,
which actually correspond 
to the Lorentz transformation in the $(i,9)$ and 
$(i,10)$ planes ($i=1,2,\cdots , 8$),
we have to take into account the phase volume analogous to 
the Faddeev-Popov determinant.
In the present case, it can be given by
the 16-dimensional volume 
spanned by the 16 vectors (\ref{zeromodes}) in the configuration space.
We denote this phase volume as ${\cal V}^{(16)}$.
Thus we arrive at the models, which describe 
the two limiting theories corresponding to
the even/odd $\nu$ sequences,
\beqa
Z_{\nu = 2 \infty}
&=& \int \left(\prod _{\mu = 1} ^8 \dd A_{\mu}\right) 
\, {\cal V}^{(16)} \, |J| ^{-1/2} \, 
\ee ^{-S_b} \,  \left| \det {\cal M}^{(8)} \right|  \  ,
\label{ZnuinftyEVEN}
\\
Z_{\nu = 2 \infty+1}
&=& \int \left(\prod _{\mu = 1} ^8 \dd A_{\mu} \right)
\,  {\cal V}^{(16)} \, |J| ^{-1/2} \, 
\ee ^{-S_b} \, \det {\cal M}^{(8)}  \ ,
\label{ZnuinftyODD}
\eeqa
where $J$ denotes the determinant of
the Hesse matrix $H_{ja,kb}$ after removing the zero-modes.
We recall that $\det {\cal M}^{(8)}$ is real but not
positive definite, which makes the two models 
{\em a priori} different. 
One can see that the exact result (\ref{trF2exact}) for 
$\langle S_b \rangle$ can be reproduced from (\ref{ZnuinftyEVEN})
and (\ref{ZnuinftyODD})
as it should, by noting that
${\cal V}^{(16)}\mapsto \lambda ^{16} {\cal V}^{(16)}$ and
$ J \mapsto \lambda ^{- 2\{2(N^2 -1) - 16\}}J  $
under a scale transformation $A_\mu \mapsto \lambda A_\mu$.

Let us consider the question whether the remaining eight-dimensional
Lorentz invariance 
of the models (\ref{ZnuinftyEVEN}) and (\ref{ZnuinftyODD})
is further broken.
For this we show that the hessian $J$ becomes zero for
configurations in $\Omega _7$ and
that the order of zero increases for configurations
in $\Omega _d$ with smaller $d$.
To quantify this statement, let us consider a configuration
with $A_1, \cdots , A_d$ being generic and $A_{d+1}, \cdots , A_8$ being
of order $\epsilon$.
We first note that each element of $H_{ja,kb}$ for such a configuration
becomes of order $\epsilon ^{(8-d)}$
due to (\ref{derivatives}).
Therefore, the non-zero eigenvalues of $H_{ja,kb}$ 
become of order $\epsilon ^{(8-d)}$.
We have checked numerically that they are not of order higher than
$\epsilon ^{(8-d)}$ generically. 
This means that 
$J$ is of order $\epsilon ^{(8-d)\, \{ 2\,(N^2 -1 ) - 16 \} }$.
On the other hand, 
degenerate configurations with smaller $d$ is suppressed by the 
entropy factor $\epsilon ^{(8-d)\, (N^2 -1)}$
coming from the integration measure.
The phase volume factor ${\cal V}^{(16)}$ gives also a suppression
of order $\epsilon ^{2 \, (8-d)}$.
Collecting all the powers of $\epsilon$, we find a
suppression of order $\epsilon ^{10 \, (8-d)}$.
This means, first of all, 
that the integrals (\ref{ZnuinftyEVEN}) and (\ref{ZnuinftyODD})
are non-singular at the degenerate configurations.
Secondly, we note that, as far as the $N$-dependent part is concerned, 
the suppression for smaller $d$ coming the entropy factor
$\epsilon ^{(8-d)\, (N^2 -1)}$ is exactly cancelled by the 
enhancement coming from $|J|^{-1/2}$.
This means that there is essentially no entropical barrier against
having smaller $d$.

So far, we have been focusing on the effect of $\Gamma ^{(i)}$ on 
the dynamics of the IIB matrix model.
The effect of $S_b$ and $\Gamma ^{(r)}$, on the other hand, has been studied
in Ref. \cite{AIKKT} by using the low-energy effective theory
obtained by the one-loop perturbation theory (\ref{perturbation}).
There, it was argued that the complicated branched-polymer interaction
among the diagonal elements $x_{i \mu}$ of the bosonic matrices $A_\mu$
might induce a collapse of the distribution of $x_{i \mu}$.
Monte Carlo simulation of the low-energy effective theory
at $\nu = 0$ \cite{oneloop} shows that the effect of 
$S_b$ and $\Gamma ^{(r)}$ is not sufficient to induce
a spontaneous breakdown of Lorentz invariance.
However, after taking into account
the effect of $\Gamma ^{(i)}$ by sending $\nu$ to infinity,
it is very plausible that the remaining eight-dimensional Lorentz invariance 
of the models (\ref{ZnuinftyEVEN}) and (\ref{ZnuinftyODD})
is further broken by the effect of $S_b$ and $\Gamma ^{(r)}$,
because there is no entropical barrier against having degenerate
configurations any more.
If this is the case, 
the vacuum at $\nu = \infty$
is given by degenerate configurations with $d < 8$.
In this regard, we recall also that 
the pfaffian $\Pf \, {\cal M}(A)$ is zero
when $\{ A_\mu \} \in \Omega _2$, 
as stated in (\ref{omega2}).
Therefore, the dimensionality $d$ of the vacuum configurations 
must be $d > 2$.

In order to check the above statements and to determine the 
dimensionality $d$ of the vacuum configurations at $\nu = \infty$,
one has to carry out the integration over the saddle-point
configurations described by
(\ref{ZnuinftyEVEN}) and (\ref{ZnuinftyODD}),
for example, by Monte Carlo simulation.
Note that the model (\ref{ZnuinftyEVEN}) is not plagued by
the sign problem any more.
%
%
The model (\ref{ZnuinftyODD}), on the other hand, 
can be studied using the configurations
generated with the model (\ref{ZnuinftyEVEN}) as
\beq
\left\langle {\cal O} \right\rangle _{\nu = 2\infty +1}
=
\frac{\left\langle {\cal O} \, \mbox{sgn}(\det {\cal M}^{(8)})
\right\rangle _{\nu = 2\infty }}
{\left\langle \mbox{sgn}(\det {\cal M}^{(8)}) 
\right\rangle _{\nu = 2\infty}}  \ .
\label{EVENtoODD}
\eeq
If the fermion determinant $\det {\cal M}^{(8)}$ 
does not have a definite sign
for dominant configurations, the denominator as well as the numerator 
of the r.h.s. in (\ref{EVENtoODD}) becomes very small, exhibiting
the sign problem.
However, if it turns out that the vacuum of the model (\ref{ZnuinftyEVEN}) 
is given by configurations with $d \le 6$, 
then due to (\ref{omega6}), we have 
$\mbox{sgn}(\det {\cal M}^{(8)})=1$ for dominant configurations.
In this case, the sign problem, which {\em a priori} exists in 
the model (\ref{ZnuinftyODD}), is solved in a sense {\em dynamically},
and the model (\ref{ZnuinftyODD}) is actually equivalent to
the model (\ref{ZnuinftyEVEN}),
since $\left\langle {\cal O} \right\rangle _{\nu = 2\infty +1}
= \left\langle {\cal O} \right\rangle _{\nu = 2\infty}$
for any observables ${\cal O}$.

\vspace*{1cm}


To summarize, 
we considered a deformation of the IIB matrix model
by introducing an integer parameter $\nu$ which couples to
$\Gamma ^{(i)}$,
the imaginary part of the effective action induced by fermions.
We studied the deformed model at $\nu = \infty$,
where the integration over the bosonic matrices is
dominated by the configurations for which $\Gamma ^{(i)}$ is stationary.
First of all,
there is still a huge configuration space
left as the saddle-point configurations which we have to integrate over.
Secondly, these saddle-point configurations have more than
two shrunken directions.
Thirdly, the more shrunken directions the configuration has, 
the more stationary $\Gamma ^{(i)}$ becomes.
This gives rise to an enhancement for configurations
with more shrunken directions,
and the effect was shown to cancel exactly the $N$-dependent entropical
barrier against having such configurations.
An intriguing feature of this enhancement is that it occurs 
exactly when the configuration becomes a lower-dimensional
hyperplane. This may be responsible for generating a 
{\em flat} space-time instead of a curved one
or a fractal one, as a result of the dynamics of the IIB matrix model.
We argued that the dimensionality $d$ of the space-time generated 
dynamically in the deformed IIB matrix model at $\nu = \infty$
is $2 < d \le 8$ and most likely $d<8$.
We derived the models which describe the integration
over the saddle-point configurations for the two {\em a priori} different
limiting theories
corresponding to the even/odd $\nu$ sequences.
Remarkably, the model with the even $\nu$ sequence 
is not plagued by the sign problem.
One can therefore study the model
by standard Monte Carlo simulation to extract the
dimensionality $d$ of the vacuum configurations.
If this turns out to be $d \le 6$,
the model with the odd $\nu$ sequence
is actually equivalent to the model with the even $\nu$ sequence.


Whether the original model ($\nu = 1$) belongs to the same phase
as $\nu = \infty$
is a nontrivial question,
which is not accessible through standard
Monte Carlo simulation due to the sign problem.
We quote, however, an example from history, where
an analogous approach was successful.
It is the strong coupling limit in the lattice formulation
of nonabelian gauge theories.
Although one should send the bare coupling constant to zero in the 
continuum limit, confinement in these theories has been clearly
demonstrated in the strong coupling limit \cite{Wilson}.
The limit also provides a qualitative 
understanding of
the spontaneous breakdown of chiral symmetry \cite{KawamotoSmit}.
%
The existence of such an approach
is indeed one of the advantages of having a nonperturbative formulation, 
which should also apply to the case of string theory.
Note, however, that
the above successes in the case of nonabelian gauge theories
rely crucially on the fact that
there is no phase transition between the strong coupling regime
and the weak coupling regime \cite{Creutz}.
An illustrative example clarifying this point is that 
confinement holds in the strong coupling limit even for
abelian gauge theories, for which there is actually a phase transition
to a deconfining phase at an intermediate coupling constant.

%
The fact that 
the model at $\nu = \infty$
still has a huge configuration space to integrate over,
which is quite peculiar to this system,
may suggest that it is in fact quite close
to the original model ($\nu =1$).
Although it is hard to justify this statement rigorously,
it would be certainly worthwhile to explore further the dynamics of 
the deformed IIB matrix model in this limit.

\vspace*{1cm}

J.\ N.\ would like to thank K.N.\ Anagnostopoulos, T.\ Hotta, T.\ Izubuchi
and A.\ Tsuchiya for collaborations at the earlier stage of this
work.
We are also grateful to J.\ Ambj\o rn, W.\ Bietenholz, 
D.\ B\"odeker, F.R.\ Klinkhamer and
N.\ Obers for valuable comments and discussions.
J.\ N.\ is supported by the Japan Society for the Promotion of
Science as a Postdoctoral Fellow for Research Abroad.  
The work of G.\ V.\ is supported by MURST (Italy) within the project
``Fisica Teorica delle Interazioni Fondamentali''.



\newpage

\begin{thebibliography}{99}


\bibitem{BFSS} T.\ Banks, W.\ Fischler, S.H.\ Shenker and L.\ Susskind,
{\em M Theory as a Matrix Model: A Conjecture,
Phys.\ Rev.} {\bf D~55} (1997) 5112 [{\tt hep-th/9610043}].

\bibitem{IKKT}N.\ Ishibashi, H.\ Kawai, Y.\ Kitazawa and A.\ Tsuchiya,
{\em A Large-$N$ Reduced Model as Superstring,
Nucl.\ Phys.} {\bf B~498} (1997) 467 [{\tt hep-th/9612115}].

\bibitem{FKKT}M.\ Fukuma, H.\ Kawai, Y.\ Kitazawa and A.\ Tsuchiya,
{\em String Field Theory from IIB Matrix Model,
Nucl.\ Phys.} {\bf B~510} (1998) 158 [{\tt hep-th/9705128}].

\bibitem{EK}
T.\ Eguchi and H.\ Kawai,
{\em Reduction of Dynamical Degrees of Freedom 
in the Large-$N$ Gauge Theory,
Phys.\ Rev.\ Lett.} {\bf 48} (1982) 1063.

\bibitem{RS}
L.\ Randall and R.\ Sundrum,
{\em An Alternative to Compactification,
Phys.\ Rev.\ Lett.} {\bf 83} (1999) 4690 [{\tt hep-th/9906064}].

\bibitem{AK}
J.\ Greensite and F.R.\ Klinkhamer,
{\em Superstring Amplitudes and Contact Interactions, Nucl.\ Phys.} {\bf B~304} (1988) 108; 
V.A.\ Kosteleck\'{y} and S.\ Samuel, {\em Spontaneous Breaking of Lorentz Symmetry in
 String Theory,  Phys.\ Rev.} {\bf D~39} (1989) 683; {\em Gravitational Phenomenology in Higher-dimensional Theories 
and Strings, Phys.\ Rev.} {\bf D~40} (1989) 1886; 
V.A.\ Kosteleck\'{y} and R.\ Potting, {\em CPT and Strings,  Nucl.\ Phys.} {\bf B~359} (1991) 545;
{\em Expectation Values, Lorentz Invariance, and CPT in the Open Bosonic String, Phys.\ Lett.} {\bf B~381} (1996) 89. V.A.\ Kosteleck\'{y}, M.J.\ Perry and R.\ Potting, {\em Off-shell Structure of the String Sigma
Model}, [{\tt hep-th/9912243}].


\bibitem{SuyamaTsuchiya}T.\ Suyama and A.\ Tsuchiya,
{\em Exact Results in $N_c =2$ IIB Matrix Model,
Prog.\ Theor.\ Phys.} {\bf 99} (1998) 321 [{\tt hep-th/9711073}].

\bibitem{AIKKT}H.\ Aoki, S.\ Iso, H.\ Kawai, Y.\ Kitazawa and T.\ Tada,
{\em Space-time Structures from IIB Matrix Model,
Prog.\ Theor.\ Phys.} {\bf 99} (1998) 713 [{\tt hep-th/9802085}].

\bibitem{KNS}W.\ Krauth, H.\ Nicolai and M.\ Staudacher,
{\em Monte Carlo Approach to M-Theory,
Phys.\ Lett.} {\bf B~431} (1998) 31 [{\tt hep-th/9803117}].

\bibitem{AABHN}J.\ Ambj\o rn, K.N.\ Anagnostopoulos, W.\ Bietenholz,
T.\ Hotta and J.\ Nishimura,
{\em Large-$N$ Dynamics of Dimensionally Reduced 4D SU($N$) Super Yang-Mills 
Theory}, [{\tt hep-th/0003208}].

\bibitem{HNT} T.\ Hotta, J.\ Nishimura and A.\ Tsuchiya, 
{\em Dynamical Aspects of Large-$N$ Reduced Models,
Nucl.\ Phys.} {\bf B~545} (1999) 543 [{\tt hep-th/9811220}].

\bibitem{sign}
W.\ Bietenholz, A.\ Pochinsky and U.-J.\ Wiese
{\em Meron-Cluster Simulation of the $\theta$ Vacuum in the 2D
O(3) Model,
Phys.\ Rev.\ Lett.} {\bf 75} (1995) 4524 [{\tt hep-lat/9505019}];\\
S.\ Chandrasekharan and U.-J.\ Wiese,
{\em Meron-Cluster Solution of a Fermion Sign Problem,
Phys.\ Rev.\ Lett.} {\bf 83} (1999) 3116 [{\tt cond-mat/9902128}].

\bibitem{oneloop}J.\ Ambj\o rn, K.N.\ Anagnostopoulos, W.\ Bietenholz,
T.\ Hotta and J.\ Nishimura,
in preparation.

\bibitem{BHN}
G.\ Bhanot, U.M.\ Heller and H.\ Neuberger, 
{\em The Quenched Eguchi-Kawai Model,
Phys.\ Lett.} {\bf 113~B} (1982) 47.


\bibitem{Wilson} K.G.\ Wilson, 
{\em Confinement of Quarks,
Phys. Rev.} {\bf D~10} (1974) 2445.

\bibitem{KawamotoSmit}
H.\ Kluberg-Stern, A.\ Morel, O.\ Napoly and B.\ Petersson,
{\em Spontaneous Chiral Symmetry Breaking for a U($N$) Gauge Theory 
on a Lattice, Nucl.\ Phys.} {\bf B~190} (1981) 504;\\
N.\ Kawamoto and J.\ Smit,
{\em Effective Lagrangian and Dynamical Symmetry Breaking in Strongly 
Coupled Lattice QCD,
Nucl.\ Phys.} {\bf B~192} (1981) 100.

\bibitem{Creutz} M.\ Creutz, 
{\em Monte Carlo Study of Quantized SU(2) Gauge Theory,
Phys.\ Rev.} {\bf D~21} (1980) 2308.

\end{thebibliography}
\end{document}